\documentclass[nofootinbib,twocolumn,showpacs,showkeys,pra,amsmath,amssymb]{revtex4}

\usepackage{graphicx}
\usepackage{hyperref}

\newcommand{\ket}[1]{|#1\rangle}
\newcommand{\bra}[1]{\langle#1|}

\newcommand{\eq}{\begin{equation}}
\newcommand{\fine}{\end{equation}}

\newcommand{\cw}{\Delta T_c}

\newcommand{\tpump}{\tau_{\text{pump}}}
\newcommand{\tphoton}{\tau_c}
\newcommand{\ii}{\text{i}}

\begin{document}


\title{
Generation of time-bin entangled photons without temporal post-selection
}
\author{Alessandro Rossi}
\homepage{http://quantumoptics.phys.uniroma1.it/}
\author{Giuseppe Vallone}
\homepage{http://quantumoptics.phys.uniroma1.it/}
\author{Francesco De Martini}
\homepage{http://quantumoptics.phys.uniroma1.it/}
\author{Paolo Mataloni}
\homepage{http://quantumoptics.phys.uniroma1.it/}
\affiliation{
Dipartimento di Fisica della ``Sapienza'' Universit\`{a} di Roma,
Roma, 00185 Italy and\\
Consorzio Nazionale Interuniversitario per le Scienze Fisiche della Materia,
Roma, 00185 Italy}

\date{\today}

\begin{abstract}
We report on the implementation of a new interferometric scheme that allows the generation of photon pairs entangled in the time-energy
degree of freedom. This scheme does not require any kind of temporal post-selection on the generated pairs and can be used even with
lasers with short coherence time.
\end{abstract}

\maketitle

\section{Introduction}\label{sec:intro}
Quantum entanglement represents the key ingredient of many quantum information processes. 
It represents a unique resource that, associated with non-classical correlations among separated quantum systems, 
can be used to perform computational \cite{01-rau-aon,01-kni-asc} and cryptographic \cite{91-eke-qua} tasks that are impossible with classical systems. 
An entangled state shared by two or more separated parties is a valuable resource for 
fundamental quantum communication protocols, such as quantum teleportation \cite{98-bos-exp,97-bou-exp}. 

In quantum optics, entanglement based on discrete variables is created by the spontaneous parametric down conversion (SPDC) process 
in a nonlinear (NL) optical crystal under excitation of a laser beam, 
either in continuous wave (CW) or pulse operation. 
By this process highly pure entangled states are produced by encoding qubits in a particular degree 
of freedom (DOF) of the photons, such as polarization \cite{95-kwi-new}, 
linear and angular momentum \cite{90-rar-exp,04-lan-mea,01-mai-ent} and energy-time \cite{89-fra-bel}. 
More recently, the possibility of spanning a large Hilbert space by encoding two photons in more than one DOF at the same time 
was demonstrated by exploiting the so-called hyperentanglement \cite{05-cin-all,05-bar-pol,05-yan-all,05-bar-gen}.  
Energy-time entanglement, also including the time-bin approach,
generally realized with fiber interferometers \cite{99-bre-pul}, is based on the
interferometric scheme proposed by J. D. Franson \cite{89-fra-bel}
which allows the creation of a superposition state of emission times.
Relevant realizations of bulk schemes of this idea were also demonstrated \cite{93-kwi-hig,94-tap-vio}.

In this paper we present a novel bulk scheme which allows the efficient creation of  
time-energy entanglement with SPDC photon pairs emitted by a Type I phase matched NL crystal.
At variance with the Franson's scheme, instead of two phase-locked interferometers (one for each photon),
this apparatus is based on a single Michelson Interferometer (MI) into which the two photons,
travelling along parallel directions, are injected.
This configuration reduces the problems of phase instability.
Furthermore, by swapping the photon modes in one of tue two MI arms,
the entangled state is automatically generated without any need of temporal post-selection. 
Furthermore, this scheme allows time-energy entanglement to be created
with any kind of lasers that also have a very short coherence time.

The paper is organized as follows.
Section \ref{sec:review} reviews the Franson's unbalanced interferometer and the conditions that must be satisfied in order to observe time-bin entanglement.
In Section \ref{sec:scheme} we describe the MI adopted in our experiment
and the generation of the entangled state .
Section \ref{sec:results} describes the results of our experiment.
Finally, in Section \ref{sec:conclusions} we give the conclusions and 
indicate some possible use of such scheme.

\section{Review of the Franson's interferometric scheme}\label{sec:review}
\begin{figure}[t]
\includegraphics[width=7cm]{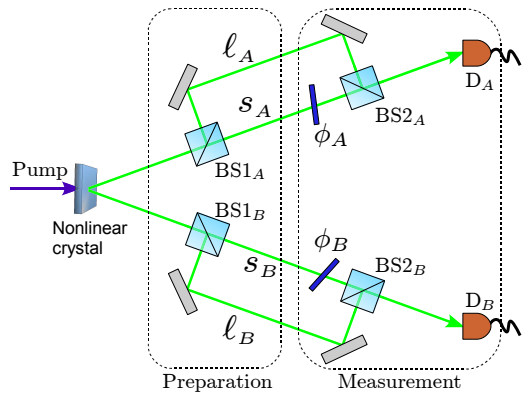}
\caption{Scheme of the Franson's interferometer. The long and short arms are labeled as $\ell$ and $s$ respectively. The thin glasses ($\phi_A$, $\phi_B$) can
be used to change the phase in the measurement.}
\label{fig:franson}
\end{figure}

Let's consider two photons, namely Alice (A) and Bob (B), emitted by a SPDC source towards two different directions. 
Each one is injected into a Mach-Zehnder (MZ) interferometer (cfr. Fig. \ref{fig:franson}).

First order interference arises if the imbalances between the MZ's long and the short arms 
\eq
\Delta x_i=\ell_i-s_i\,,\qquad i=A,B
\fine
does not exceed the single photon coherence length, $c\tphoton$. 
We can model the Franson's scheme as given by two independent sequential operations, 
corresponding respectively to a preparation and a measurement device (see Fig. \ref{fig:franson}).  
The first one is realized by the first pair of beam splitters (BS1$_i$) and generates the state
\eq
\ket{\Psi_0}=\frac12({\ket{s_A}}+{\ket{\ell_A}})({\ket{s_B}}+~{\ket{\ell_B}})\,.
\fine
The detection occurring after the second pair of beam splitters, BS2$_i$, projects $\ket{\Psi_0}$ into
the state 
\eq
\ket{\Phi_0}=\frac12({\ket{s_A}}+e^{\ii\phi_A}{\ket{\ell_A}})({\ket{s_B}}+e^{\ii\phi_B}{\ket{\ell_B}})\,.
\fine
The single counts oscillate as $\cos^2\frac{\phi_i}2$ because of single photon interference 
and the coincidence rate is expressed as
\eq\label{single_oscillation}
C(\phi)=C_0\cos^2\frac{\phi_A}2\cos^2\frac{\phi_B}2\,.
\fine

In order to create time-bin entanglement the following
conditions must be fulfilled when operating under CW pumping (we consider the two imbalances $\Delta x_A=\Delta x_B\equiv\Delta x$ equal):
 \begin{itemize}
 \item [i)]
The arm length difference must be long compared to the coherence length of photon wavepackets:
\eq\label{single}
 \Delta x> c\tphoton\,.
 \fine
This condition, avoiding single photon interference,
is typically satisfied with an imbalance $\Delta x\geq100\mu m$, 
corresponding to detect the photon in the visible range within a bandwidth of few nanometers.
  \item [ii)] The imbalance $\Delta x$ must be lower than the pump coherence length $\tpump$: 
\eq\label{couple}
\Delta x< c\tpump\,.
 \fine
This condition guarantees the coherent superposition of the ${\ket{s_A}}{\ket{s_A}}$ and ${\ket{\ell_B}}{\ket{\ell_B}}$ events.
 \item[iii)] $\Delta x$ must be long enough to discard by post-selection the events 
occurring when one photon takes the short path and the other takes the long path and vice-versa. 
These measurement outcomes correspond to the events $\ket{s_A,\ell_B}$ and $\ket{\ell_A,s_B}$. 
This latter requirement imposes the condition 
\eq\label{window}
\Delta x> c\cw\,,
\fine
where $\cw$ represents the duration of the coincidence window.
This condition imposes a strong constraint on the imbalance $\Delta x$.
 \end{itemize}

Once the above conditions are satisfied, the events $\ket{s_A,\ell_B}$ and $\ket{\ell_A,s_B}$ are distinguishable 
and may be discarded by a suitable choice of the time coincidence window, 
while $\ket{\ell_A,\ell_B}$ and $\ket{s_A,s_B}$ are indistinguishable and generates the interference.
Indeed, it is impossible to determine if the photons are emitted at time $t_0$ and both travel through the long paths 
or are emitted at time $t_0+\Delta x/c$ and both travel through the short paths.
In this configuration, the first beam splitters (BS1$_A$ and BS1$_B$) generate the state
\eq
\ket\Psi=\frac12\ket\psi\bra\psi+\frac14(\ket{s_A,\ell_B}\bra{s_A,\ell_B}+\ket{\ell_A,s_B}\bra{\ell_A,s_B})
\fine
where
\eq
\ket\psi=\frac1{\sqrt{2}}(\ket{s_A,s_B} + \ket{\ell_A,\ell_B})
\fine
The second beam splitters (BS2$_i$), together with the subsequent detection, perform the projection into the entangled state
\eq
\ket\Phi=\frac1{\sqrt{2}}(\ket{s_A,s_B} + e^{\ii(\phi_A+\phi_B)}\ket{\ell_A,\ell_B})\,.
\fine
with phase shifts $\phi_i$ realized by using two thin glass plates.

From equations \eqref{couple} and \eqref{window} it follows that the coherence time $\tpump$ of the pump beam must exceed the coincidence window
\eq\label{condition}
\tpump>\cw\;.
\fine
Typically, the minimum achievable duration of the coincidence window is $1.5 nsec$, 
hence efficient post-selection can be obtained if $\Delta x \geq 60 cm$. 
This generally imposes the use of a single longitudinal mode laser in order to satisfy conditions i), ii) and iii) in the CW regime. 
When operating with an unbalanced bulk interferometer, any length variation 
of the corresponding arms affects the stability of the entangled state, hence this configuration requires a critical stabilization 
of the phase. 
In the case of an unbalanced interferometer based on single mode fibers, 
the phase is kept stable by by temperature stabilization of the fibers.
\begin{figure}[t]
\includegraphics[width=8.5cm]{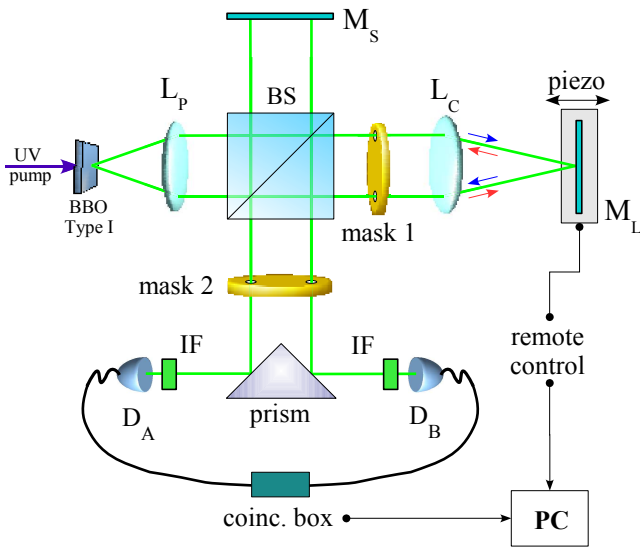}
\caption{Michelson interferometric scheme used to generate time-bin entanglement.}
\label{fig:interferometer}
\end{figure}

\section{The experiment}\label{sec:scheme} 
In the experimental apparatus  (see Fig. \ref{fig:interferometer}) SPDC photon pairs are generated 
at degenerate wavelength (wl) $\lambda= 532 nm$ by a $1 mm$ slab 
of $\beta$-Barium Borate (BBO), cut for Type I phase matching, excited by a CW single longitudinal mode laser 
(MBD-266, Coherent, coherence time $\tpump>0.1\mu sec$, $\lambda_{\text{pump}}=266nm$). 

In order to inject the two photons into a single MI, the corresponding $\bf k$ modes must be parallel.
On this purpose, the characteristic SPDC conical emission of the NL crystal is transformed 
into a cylindrical one by a spherical lens ($L_P$) whose relative distance from the BBO corresponds exactly to 
its focal length $f_L = 9.5 cm$. The lens position is carefully set by optimizing the visibility of the single photon interference.
The two photons, travelling along parallel directions, are injected into 
the input port of the MI shown in Fig. \ref{fig:interferometer}. 
In this setup the same BS is used both for the preparation of the state in the first passage of the photons (corresponding to the $BS1$'s in the Franson's scheme)
and for the final measurement in the second passage (equivalent to the $BS2$'s).

The photons' directions are carefully selected by two 2-hole masks (mask 1 and 2 in Figure \eqref{fig:interferometer}), 
the first inserted in the long arm and the second one at the MI output.
Photons can travel along the short or long path, after the partition from a symmetric ($R = T = 0.50 \pm 0.03$) beam splitter (BS). 
After reflection by mirrors $M_s$  and $M_\ell$,
each photon experiences a new BS reflection-transmission process with relative path delay $\Delta x_0=2(\ell-s)$ 
\footnote{The factor 2 with respect the Franson's scheme is due to the double passage in each arm.}. 
The BS output beams are then directed by a reflecting prism towards two single photon counting (SPCM-AQR 14) modules 
and detected within a bandwidth $\Delta\lambda=4.5 nm$, which corresponds to a single photon coherence time of $\tphoton\sim100fsec$.
The relative phase of path contributions 
$\ket{s_A,s_B}$ and $\ket{\ell_A,\ell_B}$ can be modified by a piezo transducer that finely changes the position of mirror $M_\ell$ on the $\ell$ arm.
This single MI configuration allows the phase instability problems to be minized, as said.

The key element of our setup is represented by the swapping operation between the two photons, performed in the $\ell$ arm of MI.
It is realized by the lens $L_C$ (see Figure \ref{fig:interferometer}),
aligned in such a way that mirror $M_L$ is located exactly in the focal plane of $L_C$. 
By this configuration
we can only detect coincidences if both photons travel along the same arm of the MI. In fact if, for instance, the first photon
goes through the $\ell$ path and the second follows the $s$ path (i.e. if we consider the event ${\ket{\ell}}_1{\ket{s}}_2$), 
after the second BS passage they will end on the same detector. Thus no coincidence is detected in this case. 
The same happens for the ${\ket{s}}_1{\ket{\ell}}_2$ event.
\begin{figure}[t]
\includegraphics[width=8.5cm]{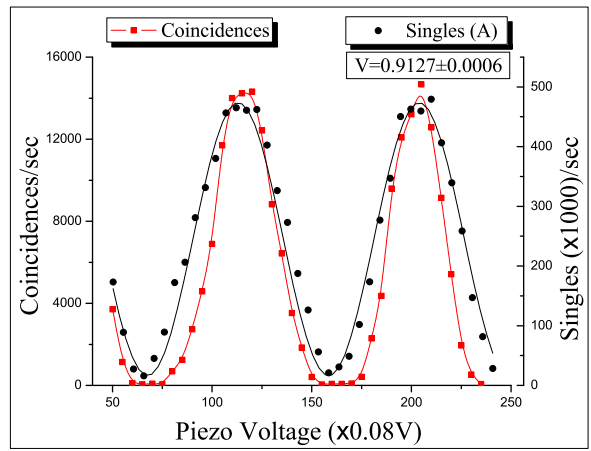}
\caption{Single (right) and coincidence (left) counts measured by our apparatus in presence of single photon interference. 
The high visibility of the single count oscillation is used to test the correct position of the lens $L_P$ . 
Coincidence counts also oscillate in agreement with eq. \eqref{single_oscillation}.}
\label{singles}
\end{figure}

By this setup it is no longer necessary to perform temporal post-selection to discard the events
${\ket{\ell}}_1{\ket{s}}_2$ and ${\ket{s}}_1{\ket{\ell}}_2$. They are simply rejected by the coincidence measurement.
In this way the conditions i) and iii) are not required anymore. The first one is not necessary because 
single photon interference cannot occur in this configuration. On the other hand, the condition iii) is not necessary because the events
${\ket{\ell}}_1{\ket{s}}_2$ and ${\ket{s}}_1{\ket{\ell}}_2$ are not discarded by the arrival time, but because of the
arrival position.
This scheme thus allows time-entanglement also with a very short coherence time of the pump. Indeed
time-entanglement is present also if the condition \eqref{condition} is not satisfied.
Moreover the the difference $\Delta x$ can be very small.
\begin{figure}[t]
\includegraphics[width=8cm]{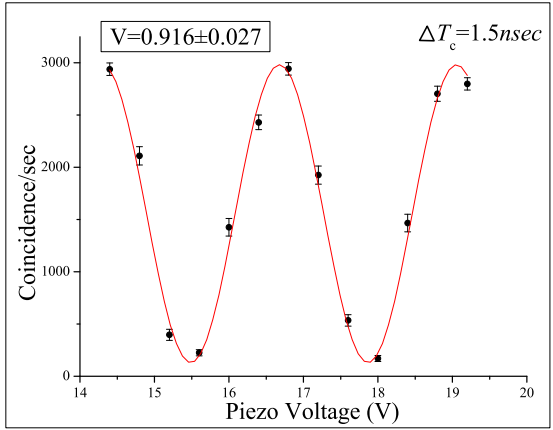}
\caption{Coincidence counts after accidental coincidence subtraction with $\cw=1.5nsec$ in the case of time-bin entanglement. 
Error bars are calculated by considering both the poissonian statistic and 
the uncertainty of the voltage applied to the piezo.}
\label{fig:1.5ns}
\end{figure}
\begin{figure}[t]
\includegraphics[width=8cm]{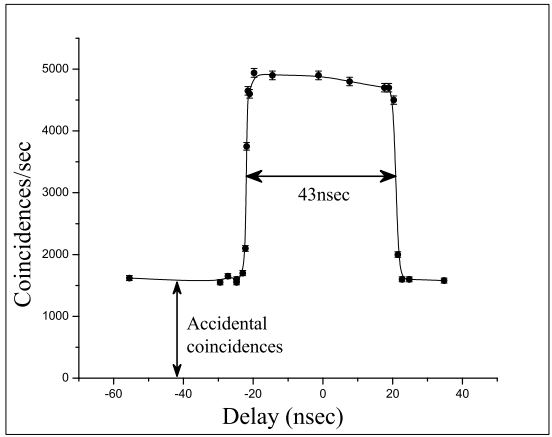}
\caption{Measurement of a coincidence window $\cw=21.5nsec$. Fixing the arrival time of the first photon, we measured the coincidences 
by varing the arrival time delay of the secon photon. The baseline ($\sim1600coinc/sec$) corresponds to the accidental coincidence contribution.
This value is in agreement with that obtained by equation \eqref{accidentals}.}
\label{fig:coinc_windows}
\end{figure}

\section{Experimental results}\label{sec:results}
Before describing the experimental results, we show the interference pattern obtained in the condition of best $L_P$ alignment. 
We used the same setup of Figure \ref{fig:interferometer} with some little
modifications on the long arm: precisely, we set its length equal to the short arm and removed the swapping lens $L_C$. 
In this way single photon interference determines a fringe pattern whose visibility is 
maximized when the two photons' directions are exactly parallel.
This is obtained when the distance between $L_P$ and the BBO crystal is equal to the lens focal lenght.
The position of $L_P$ is then carefully set by maximizing the interference.
The corresponding results are shown in Fig. \ref{singles}.

After optimization of lens $L_P$ we set the measurement apparatus in the configuration of the Figure \ref{fig:interferometer}.
Here the imbalance is $\Delta x_0=120cm$ and we do not expect any single count oscillation, while time-bin entanglement will arise.
Fig. \ref{fig:1.5ns} shows the coincidence rate as a function of the voltage applied to the piezo
for a coincidence window $\cw=1.5nsec$. 
Each datum is obtained in $1sec$ acquisition time. 
As usual, the visibility is defined as 
\eq
V=\frac{C_{\text{max}}-C_{\text{min}}}{C_{\text{max}}+C_{\text{min}}}
\fine
where $C_{\text{max(min)}}$ is the maximum (minimum) coincidence value in the oscillation pattern. Equivalently,
it is defined as the parameter $V$ in the fitting function $C(x)=c_0(1+V\cos[\omega(x-x_0)])$, with $x$ corresponding to the piezo voltage.
The high visibility value of the $V=0.916\pm0.027$ clearly demonstrates time-bin entanglement. In fact if the events
$\ket{s_A,\ell_B}$ and $\ket{\ell_A,s_B}$ were not discarded the visibility could not be larger than $0.5$.
It is worth noting that the small value of $\cw$ would allow these events to be discarded even in the case of a standard Franson's scheme.
\begin{figure}[t]
\includegraphics[width=8cm]{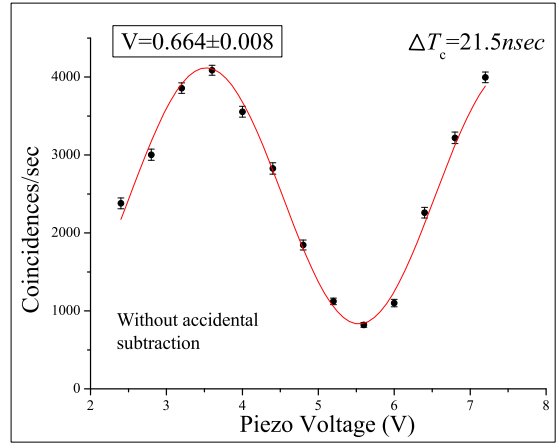}
\caption{Raw coincidence counts (without accidental coincidence subtraction) with coincidence window $\cw=21.5nsec$.}
\label{fig:21ns}
\end{figure}
\begin{figure}[t]
\includegraphics[width=8cm]{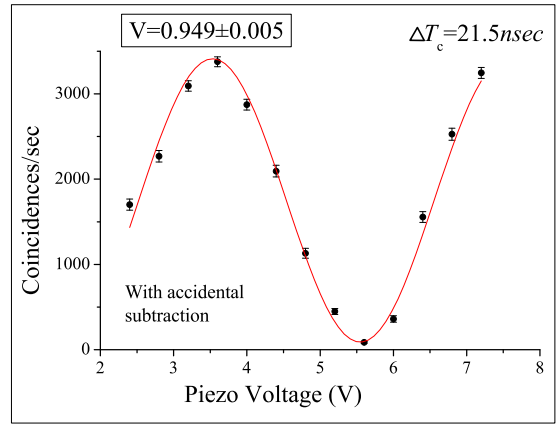}
\caption{Coincidence counts with $21.5nsec$ coincidence window after accidental subtraction.}
\label{fig:21ns_noacc}
\end{figure}

\begin{figure*}[t]
\includegraphics[width=14cm]{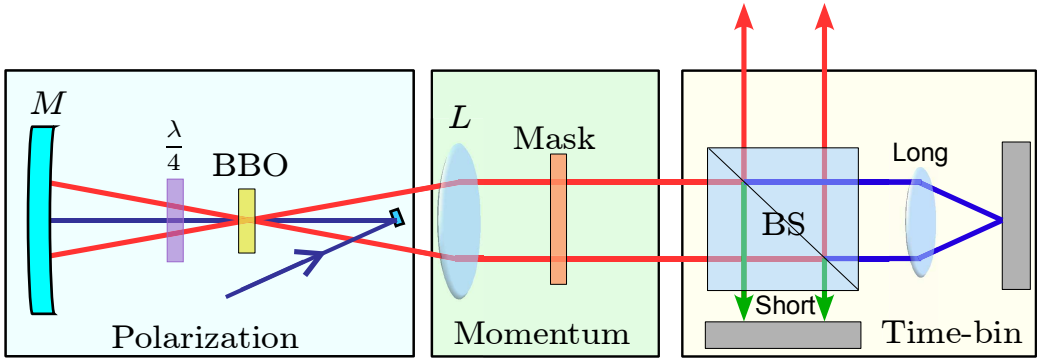}
\caption{Scheme for the generation of two photon polarization/momentum/time-bin entangled states.}
\label{fig:time-bin}
\end{figure*}
Let's now describe the case of a much longer coincidence window, $\cw=21.5nsec$.
The value of $\cw$ was measured by observing the coincidence pattern as a function of the relative temporal delay between the two photons 
(see Fig. \ref{fig:coinc_windows}).
Precisely, we fix the arrival time of the first photon and measure the coincidence number by varying the arrival time $\delta t$ of the second photon.
Since the coincidence box detects coincidences regardless of which photon comes first, we detect real coincidences if $-21.5nsec\leq\delta t\leq21.5nsec$.
In this condition, by using the standard Franson's scheme the imbalance $\Delta x_0$ would not allow time-bin entanglement
because the condition \eqref{window} would be violated.
In our scheme equation \eqref{window} is not necessary anymore; here we can generate time-bin entanglement even if
the imbalance $\Delta x_0$ is shorter than the coincidence window $\cw$. 
In Fig. \ref{fig:21ns} we show the coincidence pattern obtained with $\cw=21.5nsec$. We report here the raw data, i.e. without
any accidental coincidence subtraction. As expected, even if the visibility is quite low $V=0.664\pm0.008$, it is remarkable 
that this value is larger than the classical limit $0.5$. In this condition the standard
Franson's scheme would not allow a visibility~$>0.5$,
due to the impossibility of discarding the events $\ket{s_A,\ell_B}$ and $\ket{\ell_A,s_B}$. 
We also report in Fig. \ref{fig:21ns_noacc} the experimental results of Fig. \ref{fig:21ns} after accidental coincidence subtraction. 
These are estimated by using the expression 
\eq\label{accidentals}
C_{acc}=2S_AS_B\cw\,,
\fine
where $S_A$ and $S_B$ are the single counts measured
at detectors $D_A$ and $D_B$ respectively.
The new visibility is now $V=0.949\pm0.005$ which is comparable within the errors  with that obtained with $\cw=1.5nsec$.

\section{Conclusions}\label{sec:conclusions}

In this paper we have described a novel interferometric scheme to create time-bin entanglement based on two photon states. 
It consists of a Michelson interferometer in which two degenerate photons, created by spontaneous parametric down conversion 
in a Type I phase matched NL crystal and travelling along parallel directions, are injected.
Compared with the standard Franson's scheme, which adopts two different unbalanced interferometers and creates time-bin entanglement 
by temporal post-selection, our scheme presents significant advantages:
\begin{itemize}
 \item Since it is based on a common interferometer for the two photons, it avoids most of the problems related to phase instabilities. 
\item The scheme avoids the use of temporal post-selection to generate the entanglement. Indeed, by using a simple lens in one of the interferometer arms, 
it operates under the spatial swapping of the two photon $\bf k$-modes. 
\item Under CW operation, this scheme can work with any kind of laser (even with a short coherence time) used to pump the parametric process.
\end{itemize}
The experimental results obtained by this scheme indicate that it represents a possible solution for several applications of 
quantum communication and quantum computation. In particular, it can be used to increase the dimension of a multi-degree of freedom two photon state. 
For instance, it can be used to add two more qubits by time-energy entanglement to a four qubit polarization-momentum hyperentangled state \cite{08-val-act,08-val-one}.
By referring to the parametric source sketched in Figure \ref{fig:time-bin}, 
at the beginning, polarization entanglement is generated by double passage of the pump beam and of the photon pair through a Type I BBO crystal, 
and a $\lambda/4$ wave plate after reflection from a spherical mirror M. 
Then, linear momentum entanglement is obtained by properly selecting (for instance with a mask) 
four $\bf k$ modes of the SPDC conical emission of the Type I crystal. In this way polarization-momentum hyperentanglement is generated 
(see \cite{05-bar-pol} for a detailed description of the hyperentangled source). This state corresponds to a two-photon four-qubit state. 
By injecting the four modes into the MI described in this paper we will add time-bin entanglement and allow each photon to encode 
three qubits in the different DOF's. This will generate the polarization/momentum/time-bin entanglement of two photons.
This experiment is at the moment under investigation.

\begin{acknowledgments}
Thanks to Gaia Donati for the contribution given in realizing the experiment. This work was supported by the 
PRIN 2005 of MIUR (Ministero dell'Universi\`a e della Ricerca), Italy.   
\end{acknowledgments}

\end{document}